\documentclass{ismdproc}
\usepackage{subfigure}

\begin{document}
\title{ADHYDRO --- hydrodynamics-like model \\ for highly anisotropic systems}

\author{{\slshape Wojciech Florkowski}  \\[1ex]
Institute of Physics, Jan Kochanowski University, \\ PL-25406~Kielce, Poland, and \\
The H. Niewodnicza\'nski Institute of Nuclear Physics, Polish Academy of Sciences, \\ PL-31342 Krak\'ow, Poland }

\contribID{xy}  
\confID{yz}
\acronym{ISMD2010}
\doi            

\maketitle

\begin{abstract}
A new framework (ADHYDRO) is presented that has been designed for description of highly-anisotropic systems produced at the very early stages of relativistic heavy-ion collisions. The structure of ADHYDRO is very much similar to perfect-fluid hydrodynamics, however, ADHYDRO includes dissipation effects. Dissipation is defined by the form of the entropy source that depends on the pressure anisotropy and vanishes for the isotropic systems. With a simple ansatz for the entropy source, that obeys general physical requirements, we obtain the system of equations describing isotropization of the anisotropic systems.  
\end{abstract}

\section{Introduction}

The soft hadronic data obtained in the heavy-ion experiments at RHIC, in particular the measured values of the elliptic flow coefficient $v_2$,  are most often interpreted as the evidence for a very fast equilibration of the produced matter (most likely within a fraction of 1~fm/c) and for its perfect-fluid behavior  \cite{Kolb:2003dz,Huovinen:2003fa}. Such very fast equilibration is naturally explained by the assumption that the produced matter is a strongly coupled quark-gluon plasma (sQGP) \cite{Gyulassy:2004zy,Shuryak:2004cy,Blaizot:2007sw}. There are, however, other explanations that assume that the plasma is weakly interacting. In this case the plasma instabilities lead to the fast isotropization of matter, an effect that helps to achieve the full equilibration in a short time \cite{Mrowczynski:2005ki}. 

Several recent calculations have demonstrated that the large values of the elliptic flow measured at RHIC  may be successfully reproduced in the models that do not assume fast equilibration. In Ref. \cite{Broniowski:2008qk} the stage described by the perfect-fluid hydrodynamics was preceded by the free streaming of partons (see also \cite{Gyulassy:2007zz,Akkelin:2009nz}), while in Refs. \cite{Bialas:2007gn,Chojnacki:2007fi} the authors assume  that only transverse degrees of freedom are thermalized. The approach towards the full equilibration was discussed in this context in Refs. \cite{Ryblewski:2009hm,Florkowski:2009wb,Ryblewski:2010tn}. Such results indicate that the assumption of the fast equilibration/isotropization might be relaxed (see, however, the very recent results concerning the directed flow  \cite{Bozek:2010aj,Bozek:2010cf}, which suggest that thermalization may be indeed very fast). 

We note that the idea of practically instantaneous equilibration seems to contradict the results of the microscopic models of heavy-ion collisions. These models typically use the concepts of color strings or color-flux tubes. The matter produced by strings is highly anisotropic; the transverse pressure is usually much larger than the longitudinal pressure (the transverse and longitudinal directions are defined with respect to the beam). Similar situation takes place in the Color Glass Condensate (CGC) approach where the longitudinal momentum distribution is much narrower than the transverse one \cite{Kovner:1995ja,Bjoraker:2000cf,El:2007vg}. 

In this talk we describe briefly the framework of ADHYDRO --- highly-anisotropic and strongly-dissipative hydrodynamics \cite{Florkowski:2010cf,Ryblewski:2010bs}. This framework has been designed for modeling of very early stages of heavy-ion collisions, where high anisotropy of the produced system is expected. In Ref. \cite{Florkowski:2010cf} ADHYDRO was used to describe boost-invariant, one-dimensional  systems, whereas in Ref. \cite{Ryblewski:2010bs} ADHYDRO was applied in non-boost-invariant situations. Sensitivity of the model to various modifications of its assumptions has been analyzed recently in \cite{Ryblewski:2010ch}.

The proposed model has a structure that is very much similar to the perfect-fluid hydrodynamics. The main two differences are connected with: i) the possibility that the longitudinal and transverse pressures are different, and ii) the possibility of entropy production. We note that by relaxing the assumption about the isentropic flow, we have generalized our previous formulations of anisotropic (magneto)hydrodynamics presented earlier in Refs. \cite{Florkowski:2008ag,Florkowski:2009sw}. 

It is important to emphasize that deviations from equilibrium are naturally described in the framework of viscous (Israel-Stewart) hydrodynamics, for a recent review see \cite{Heinz:2009xj}. However, the region of the applicability of viscous hydrodynamics extends to the systems that are close to equilibrium (i.e., with small anisotropy). Thus, from the formal point of view, the use of viscous hydrodynamics in the description of very early stages of the heavy-ion collisions may be inadequate, since the strong reduction of the initial longitudinal pressure leads to significant deviations from equilibrium. 

One may show that for small pressure anisotropies and purely longitudinal expansion ADHYDRO agrees with the 2nd order Israel-Stewart viscous hydrodynamics (we note that longitudinal expansion dominates at the early stages of heavy-ion collisions). On the other hand, for very large anisotropies ADHYDRO should be treated as a hydrodynamics-like model, that has no direct relation to the Israel-Stewart theory. Certainly, if the anisotropies are very large, one should use the kinetic theory (which has its own well known limitations) or consider specific models of the collisions. ADHYDRO is a proposal for such modeling. Other possibility is to use the method presented in Refs. \cite{Bozek:2010aj,Bozek:2010cf,Bozek:2007di}, where an exponentially decreasing time dependence of the difference $P_\perp-P_\parallel$ is assumed.

A work along similar lines have been recently done by Martinez and Strickland \cite{Martinez:2010sc,Martinez:2010sd}, who derive analogous equations from the Boltzmann equation with the collision term treated in the relaxation time approximation. Instead of the usual expansion around the isotropic one-particle distribution, they consider an expansion around an anisotropic distribution.

\section{Basic equations}

The ADHYDRO model is defined by the following form of the energy-momentum tensor  \cite{Florkowski:2008ag,Florkowski:2009sw}
\begin{eqnarray}
T^{\mu \nu} &=& \left( \varepsilon  + P_\perp\right) U^{\mu}U^{\nu} 
- P_\perp \, g^{\mu\nu} - (P_\perp - P_\parallel) V^{\mu}V^{\nu},
\label{Tmunudec}
\end{eqnarray}
where $\varepsilon$, $P_\perp$, and $P_\parallel$ are the energy density, transverse pressure, and longitudinal pressure, respectively. In the case \mbox{$P_\perp=P_\parallel=P$} the form of the perfect-fluid energy-momentum tensor is reproduced. The four-vector $U^\mu$ describes the standard hydrodynamic flow
\begin{equation}
U^\mu = \gamma (1, v_x, v_y, v_z), \quad \gamma = (1-v^2)^{-1/2},
\label{Umu}
\end{equation}
and $V^\mu$ defines the longitudinal direction (corresponding to the collision axis)
\begin{equation}
V^\mu = \gamma_z (v_z, 0, 0, 1), \quad \gamma_z = (1-v_z^2)^{-1/2}.
\label{Vmu}
\end{equation}
The four-vectors $U^\mu$ and $V^\mu$ satisfy the following normalization conditions: $U^2 = 1$, \mbox{$V^2 = -1$}, and $U \cdot V = 0$. In the local-rest-frame of the fluid element, where ${\bf v}=0$, the four-vectors $U^\mu$ and $V^\mu$ are reduced to the simple forms, $U^\mu = (1,0,0,0)$, $V^\mu = (0,0,0,1)$, and the energy-momentum tensor has the following diagonal structure
\begin{equation}
T^{\mu \nu} =  \left(
\begin{array}{cccc}
\varepsilon & 0 & 0 & 0 \\
0 & P_\perp & 0 & 0 \\
0 & 0 & P_\perp & 0 \\
0 & 0 & 0 & P_\parallel
\end{array} \right).
\label{Tmunuarray}
\end{equation}
In addition to the energy-momentum tensor (\ref{Tmunudec}) we introduce the entropy flux
\begin{eqnarray}
\sigma^{\mu} &=& \sigma U^{\mu},
\label{smudec}
\end{eqnarray}
where $\sigma$ is the entropy density. The quantities $\varepsilon$ and  $\sigma$ are functions of the two pressures, $P_\perp$ and $P_\parallel$. In particular, for massless partons $T^\mu_{\,\,\mu} = 0$  and $\varepsilon = 2 P_\perp + P_\parallel$. 

\begin{figure}[t]
\begin{center}
\subfigure{\includegraphics[angle=0,width=0.56\textwidth]{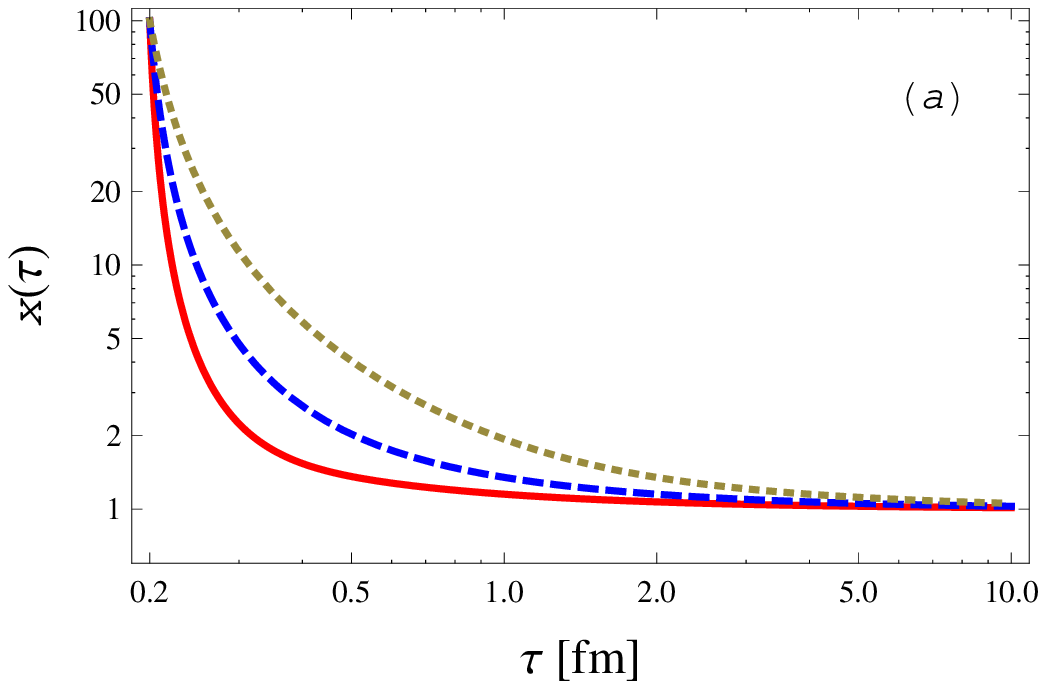}}  \\
\subfigure{\includegraphics[angle=0,width=0.56\textwidth]{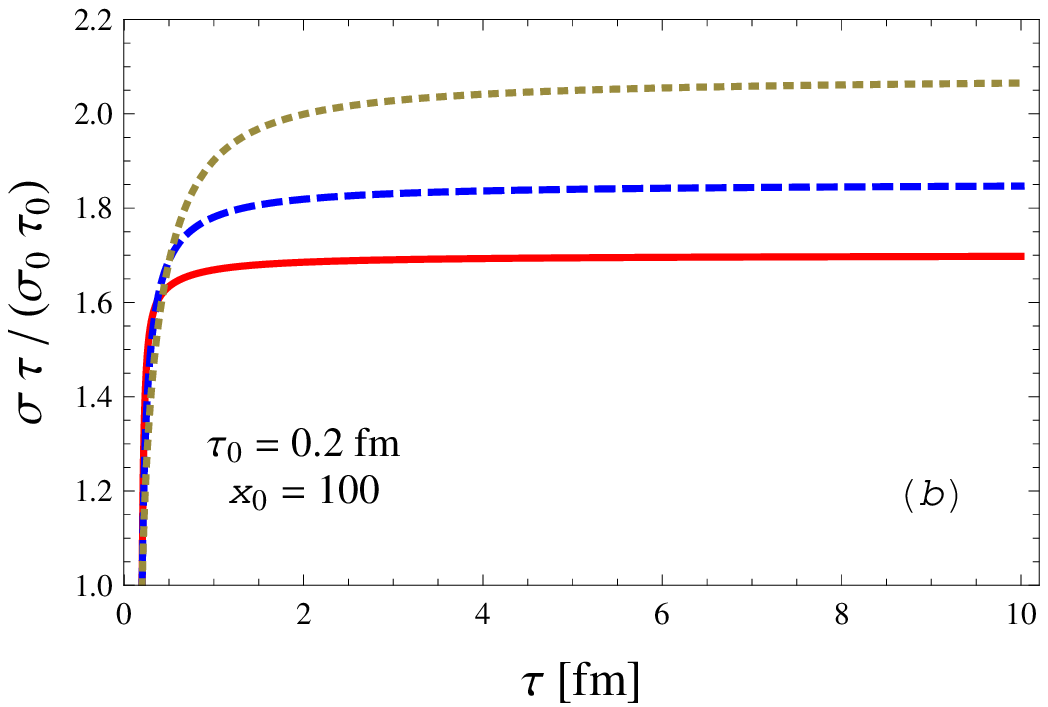}}   \\
\subfigure{\includegraphics[angle=0,width=0.56\textwidth]{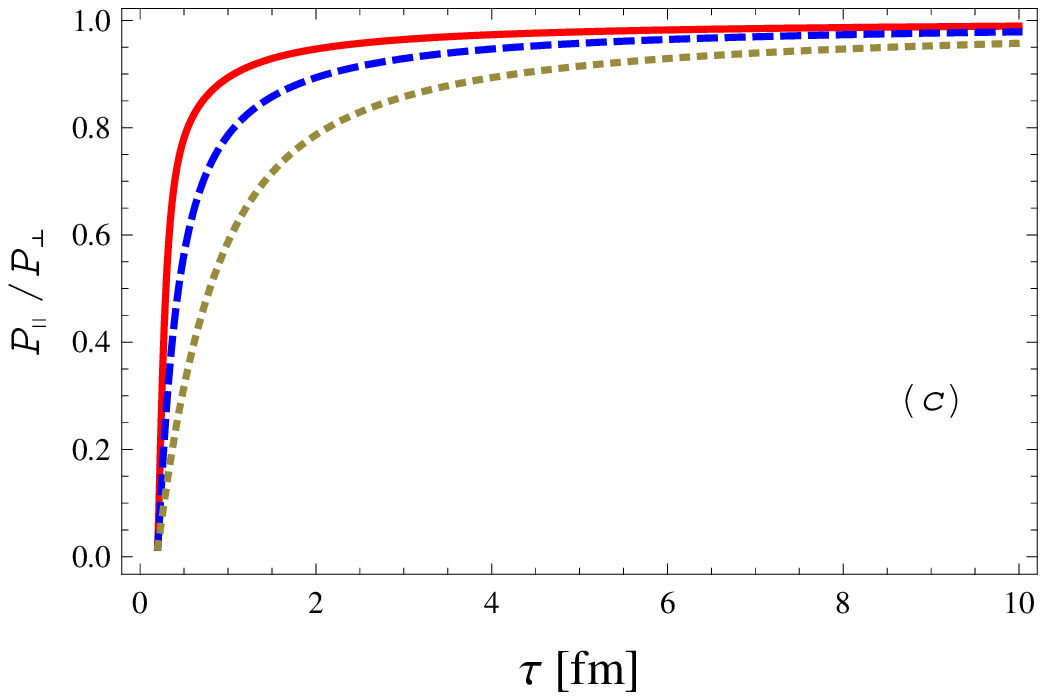}} 
\end{center}
\caption{{\bf (a)} Time dependence of the asymmetry parameter $x$ for three different choices of the relaxation time: \mbox{$\tau_{\rm eq}$ =  0.25 fm} (solid line), \mbox{$\tau_{\rm eq}$ =  0.5 fm} (dashed line), and \mbox{$\tau_{\rm eq}$ =  1.0 fm} (dotted line).  {\bf (b)} Entropy density divided by the corresponding values obtained in the Bjorken model. {\bf (c)} Ratio of the longitudinal and transverse pressures shown as a function of the proper time. All results are obtained with the initial asymmetry $x_0=100$.}
\label{fig:XBjPLPT}
\end{figure}

The space-time dynamics of the system is described by the following equations
\begin{eqnarray}
\partial_\mu T^{\mu \nu} &=& 0, \label{enmomcon} \\
\partial_\mu \sigma^{\mu} &=& \Sigma. \label{engrow}
\end{eqnarray}
Equation  (\ref{enmomcon}) expresses the energy-momentum conservation, while Eq. (\ref{engrow}) describes the entropy production. The form of the entropy source $\Sigma$ determines the dynamics of the anisotropic system. In Ref. \cite{Florkowski:2010cf} we proposed the following ansatz
\begin{equation}
\Sigma = \frac{(1-\sqrt{x})^{2}}{\sqrt{x}}\frac{\sigma}{\tau_{\rm eq}},
\label{en1}
\end{equation}
where $x$ is the parameter describing the anisotropy of the momentum distribution function of partons and $\tau_{\rm eq}$ is a time-scale parameter. One may check that \mbox{$x \approx (P_\perp/P_\parallel)^{\,4/3}$} \cite{Florkowski:2010cf}.

The expression on the right-hand-side of Eq. (\ref{en1}) has several appealing features: i) it is positive, as required by the second law of thermodynamics, ii) it has a correct dimension, iii) it vanishes in equilibrium, where $x=1$. Moreover, for small deviations from equilibrium, where $|x-1| \ll 1$, we find
\begin{equation}
\Sigma  \approx \frac{(x-1)^{2}}{4 \tau_{\rm eq}} \sigma.
\label{en1exp}
\end{equation}
The quadratic dependence displayed in (\ref{en1exp}) is characteristic for the 2nd order viscous hydrodynamics where the entropy production is proportional to the viscous stress squared, $\Sigma \propto \Pi^2$~\cite{Muronga:2003ta}. As is shown in \cite{Martinez:2010sc}, $\Pi \propto x-1$ in our case.

\section{Results for boost-invariant systems}

If the functions $\sigma(P_\perp,P_\parallel)$ and $\Sigma(P_\perp,P_\parallel)$ are specified, Eqs. (\ref{enmomcon}) and (\ref{engrow}) form a closed system of 5 equations for 5 unknown functions: three components of the fluid velocity ${\bf v}$, the transverse pressure $P_\perp$, and the longitudinal pressure $P_\parallel$. In the case of the longitudinal and boost-invariant evolution only two functions are independent. Following Ref. \cite{Florkowski:2010cf}, we express all thermodynamics-like quantities in terms of $x$ and $\sigma$ rather than in terms of $P_\perp$ and $P_\parallel$. The four-vectors $U^\mu$ and $V^\mu$ are defined by the expressions
\begin{eqnarray}
U^{\mu} = (\cosh \eta , 0 , 0 ,\sinh \eta), \label{fig:Umu} \\
V^{\mu} = (\sinh \eta , 0 , 0 ,\cosh \eta),\label{fig:Vmu}
\end{eqnarray}
where $\eta$ is the space-time rapidity, $\eta = 1/2 \ln (t+z)/(t-z)$. Substituting Eqs. (\ref{fig:Umu}) and (\ref{fig:Vmu}) in Eqs. (\ref{enmomcon}) and (\ref{engrow}) we obtain
\begin{eqnarray}
&& \frac{d \varepsilon(\sigma,x)}{d \tau} = -\frac{\varepsilon(\sigma,x) + P_\parallel(\sigma,x) }{\tau},
\label{eq1binv} \\
&& \frac{d \sigma}{d \tau} + \frac{\sigma}{\tau} = \Sigma(\sigma,x),
\label{eq3binv}
\end{eqnarray}
where $\tau = \sqrt{t^2 - z^2}$ is the proper time. The explicit forms of the functions $\varepsilon(\sigma,x)$ and $P_\parallel(\sigma,x)$, determined by the microscopic phase-space distribution function, have been given in \cite{Florkowski:2010cf}.

We solve Eqs. (\ref{eq1binv}) and (\ref{eq3binv})  numerically with $x=x_0$ set at \mbox{$\tau = \tau_0$ = 0.2 fm}. The results of the microscopic models suggest that $P_\parallel \ll P_\perp$ at the early stages of the collisions, hence we consider the case $x_0=100$. The time evolution is studied in the time interval: \mbox{0.2 fm $\leq \tau \leq$ 10 fm}. In Fig.~\ref{fig:XBjPLPT} we show the time dependence of various physical quantities obtained for three different choices of the relaxation time: \mbox{$\tau_{\rm eq}$ =  0.25 fm} (solid line), \mbox{$\tau_{\rm eq}$ =  0.5 fm} (dashed line), and \mbox{$\tau_{\rm eq}$ =  1.0 fm} (dotted line). Figure \ref{fig:XBjPLPT} (a) shows the time dependence of the asymmetry parameter $x$.  We observe fast changes of $x$ at the beginning of the evolution, which are followed by more moderate changes at later times. The behavior shown in  Fig.~\ref{fig:XBjPLPT} (a) indicates that $x \approx 1$ for $\tau \geq 2 \tau_{\rm eq}$. This is a desired effect showing that the system approaches the state of local equilibrium. 

In Fig.~\ref{fig:XBjPLPT} (b) we compare the time evolution of the entropy density obtained from Eq. (\ref{eq3binv})  with the Bjorken solution $\sigma_{\rm Bj} = \sigma_0 \tau_0/\tau$. Here $\sigma_0$ is the initial value of the entropy density. We note that the specific value of $\sigma_0$ is irrelevant for our analysis. The amount of the entropy produced in the regime described by the anisotropic hydrodynamics depends in our case on the relaxation time. For \mbox{$\tau_{\rm eq}$ =  0.25, 0.5, and 1.0 fm} the entropy increases by about 70\%, 85\% and 105\%, respectively. For $\tau \gg \tau_{\rm eq}$ the ratio $(\sigma \tau)/(\sigma_0 \tau_0)$ saturates indicating that the flow attains the form of the Bjorken flow. This  behavior shows explicitly that our framework may be used to model the transition between the highly anisotropic initial behavior and the perfect-fluid stage. In Fig.~\ref{fig:XBjPLPT} (c) we show the time dependence of the ratio of the longitudinal and transverse pressure. Again, we show three different time evolutions corresponding to three different relaxation times. For $\tau \gg \tau_{\rm eq}$ the ratio approaches unity and the two pressures become equal. 

\section{Summary}
\label{sect:con}

We have introduced the new framework of highly-anisotropic hydrodynamics with strong dissipation. The effects of the dissipation are introduced by the special form of the internal entropy source.  The source depends on the pressure anisotropy and vanishes for the isotropic systems to guarantee that the perfect-fluid behavior is reproduced for the locally equilibrated system.

Our numerical results have been presented for a simple one-dimensional system. Nevertheless, the proposed formalism is general and may be applied to more complicated 2+1 and 3+1 situations. In addition, different forms of the entropy source inspired by different microscopic mechanisms may be analyzed.

\medskip

The results presented in this paper were obtained in collaboration with Radoslaw Ryblewski. This work was supported in part by the Polish Ministry of Science and Higher Education, grant No. N  N202 263438.



\begin{footnotesize}

\end{footnotesize}



\begin{thebibliography}{99}

\bibitem{Kolb:2003dz}
P.~F. Kolb, U.~Heinz, in Quark-Gluon Plasma 3, edited by R.C. Hwa and X.-N.
  Wang (World Scientific, Singapore, 2004), p. 634.

\bibitem{Huovinen:2003fa}
P.~Huovinen, in Quark-Gluon Plasma 3, edited by R.C. Hwa and X.-N. Wang (World
  Scientific, Singapore, 2004), p. 600.

\bibitem{Gyulassy:2004zy}
M. Gyulassy, L. McLerran, Nucl. Phys. {\bf A750} (2005) 30.

\bibitem{Shuryak:2004cy}
E. V. Shuryak, Nucl. Phys. {\bf A750} (2005) 64.

\bibitem{Blaizot:2007sw}
J.-P. Blaizot,  J. Phys. {\bf G34} (2007) S243. 

\bibitem{Mrowczynski:2005ki}
S.~Mrowczynski, Acta Phys. Polon., {\bf B37} (2006) 427.

\bibitem{Broniowski:2008qk}
W.~Broniowski, W.~Florkowski, M.~Chojnacki, A.~Kisiel, Phys. Rev., {\bf C80} (2009) 034902.

\bibitem{Gyulassy:2007zz}
M.~Gyulassy, Y.~M. Sinyukov, I.~Karpenko, A.~V. Nazarenko, Braz. J. Phys., {\bf
  37} (2007) 1031.

\bibitem{Akkelin:2009nz}
S.~V. Akkelin, Y.~M. Sinyukov, Phys. Rev., {\bf C81} (2010) 064901.

\bibitem{Bialas:2007gn}
A.~Bialas, M.~Chojnacki, W.~Florkowski, Phys. Lett., {\bf B661} (2008) 325.

\bibitem{Chojnacki:2007fi}
M.~Chojnacki, W.~Florkowski, Acta Phys. Pol., {\bf B39} (2008) 721.

\bibitem{Ryblewski:2009hm}
R.~Ryblewski, W.~Florkowski, Acta Phys. Pol. (Proceedings Supplement), {\bf B3}
  (2010) 557.
  
\bibitem{Florkowski:2009wb}
W. Florkowski, R. Ryblewski, J. Phys. {\bf G37} (2010) 094023. 

\bibitem{Ryblewski:2010tn}
R.~Ryblewski, W.~Florkowski, Phys. Rev. {\bf C82} (2010) 024903.

\bibitem{Bozek:2010aj}
P. Bozek, I. Wyskiel-Piekarska, arXiv:1009.0701 [nucl-th]. 

\bibitem{Bozek:2010cf} 
P. Bozek, I. Wyskiel-Piekarska, arXiv:1011.6210 [nucl-th]. 

\bibitem{Kovner:1995ja}
A.~Kovner, L.~D. McLerran, H.~Weigert, Phys. Rev., {\bf D52} (1995) 6231.

\bibitem{Bjoraker:2000cf}
J.~Bjoraker, R.~Venugopalan, Phys. Rev., {\bf C63} (2001) 024609.

\bibitem{El:2007vg}
A.~El, Z.~Xu, C.~Greiner, Nucl. Phys., {\bf A806} (2008) 287.

\bibitem{Florkowski:2010cf}
W.~Florkowski, R.~Ryblewski, arXiv:1007.0130 [nucl-th].

\bibitem{Ryblewski:2010bs}
R.~Ryblewski, F.~Florkowski, arXiv:1007.4662 [nucl-th].

\bibitem{Ryblewski:2010ch}
R. Ryblewski, W. Florkowski, arXiv:1011.6213 [nucl-th].

\bibitem{Florkowski:2008ag}
W.~Florkowski, Phys. Lett., {\bf B668} (2008) 32.

\bibitem{Florkowski:2009sw}
W.~Florkowski, R.~Ryblewski, Acta Phys. Polon., {\bf B40} (2009) 2843.

\bibitem{Heinz:2009xj}
U.~W. Heinz, arXiv:0901.4355 [nucl-th].

\bibitem{Bozek:2007di}
P.~Bozek, Acta Phys. Polon., {\bf B39} (2008) 1375.

\bibitem{Martinez:2010sc}
M.~Martinez, M.~Strickland, Nucl. Phys. {\bf A848} (2010) 183.

\bibitem{Martinez:2010sd}
M.~Martinez, M.~Strickland, arXiv:1011.3056 [nucl-th].

\bibitem{Muronga:2003ta}
A. Muronga, Phys. Rev. {\bf C69} (2004) 034903. 

\end{thebibliography}
\end{document}